\documentclass[trackchanges]{aastex701}

\usepackage{graphicx} 
\usepackage{amsmath}
\usepackage{subcaption}
\usepackage{threeparttable}

\begin{document}

\title{Experimental Demonstration of Double-Fourier Interferometer Imaging for Future Space-Based Far-Infrared Observatories}

\author[0009-0000-3882-9242]{Jeremy P. Scott}
\affiliation{Department of Astronomy and Steward Observatory, The University of Arizona, 933 N Cherry Ave, Tucson, AZ 85719, USA}
\affiliation{Department of Physics and Astronomy, Institute for Space Imaging Science, University of Lethbridge, 4401 University Dr W, Lethbridge, AB T1K 3M4, Canada}
\email{jscott25@arizona.edu}

\author[0000-0002-9941-2077]{Locke D. Spencer}
\affiliation{Department of Physics and Astronomy, Institute for Space Imaging Science, University of Lethbridge, 4401 University Dr W, Lethbridge, AB T1K 3M4, Canada}
\email{locke.spencer@uleth.ca}

\author[0000-0001-5828-5246]{Chris Benson}
\affiliation{Department of Physics and Astronomy, Astronomy Instrumentation Group, Cardiff University, 5 The Parade, Cardiff, CF24 3AA, UK}
\affiliation{Department of Physics and Astronomy, Institute for Space Imaging Science, University of Lethbridge, 4401 University Dr W, Lethbridge, AB T1K 3M4, Canada}
\email{BensonC@cardiff.ac.uk}

\author[0000-0002-4692-7798]{David Naylor}
\affiliation{Department of Physics and Astronomy, Institute for Space Imaging Science, University of Lethbridge, 4401 University Dr W, Lethbridge, AB T1K 3M4, Canada}
\email{naylor@uleth.ca}

\author[0000-0001-9701-5660]{Berke Ricketti}
\affiliation{Disruptive Space Technology Centre, RAL Space, STFC, Fermi Ave, Harwell Campus, Didcot OX11 0QX, UK}
\affiliation{Department of Physics and Astronomy, Institute for Space Imaging Science, University of Lethbridge, 4401 University Dr W, Lethbridge, AB T1K 3M4, Canada}
\email{berke.ricketti@stfc.ac.uk}

\author[0009-0005-2514-6799]{Dinula De Silva}
\affiliation{Department of Physics and Astronomy, Institute for Space Imaging Science, University of Lethbridge, 4401 University Dr W, Lethbridge, AB T1K 3M4, Canada}
\email{dinulasd@gmail.com}

\author[0000-0002-0580-0854]{Vince Weiler}
\affiliation{Department of Physics and Astronomy, Institute for Space Imaging Science, University of Lethbridge, 4401 University Dr W, Lethbridge, AB T1K 3M4, Canada}
\email{vince.weiler@uleth.ca}


\begin{abstract}
Far-infrared (FIR) astronomy is crucial for observing star formation, active galactic nuclei, and protoplanetary systems, but its effectiveness is severely limited by poor diffraction-limited spatial resolution. To overcome this limitation, a double-Fourier interferometer (DFI), which couples a spatial interferometer with a Fourier transform spectrometer, has been proposed to deliver high-resolution broadband spectral imaging. Despite various conceptual and prototype studies, a comprehensive demonstration of DFI capabilities has been lacking. In this study, we present the validation of key DFI observing techniques using a laboratory testbed operating in the 0.3 to 2.1 THz optical bandwidth. We detail the instrument's ideal response alongside necessary calibrations for intensity imbalances, non-linear phase errors, and phase-center offsets. Our experimental results successfully demonstrate aperture synthesis for both one- and two-dimensional targets, accurately recovering source geometries. Furthermore, we prove the instrument's sensitivity to spatially dependent spectral variations by utilizing a low-pass optical filter on extended sources. Finally, we validate wide-field imaging capabilities by combining data from independent pixels in a detector array, each sensitive to different regions of the source plane. The experimental data show significant agreement with theoretical predictions, with minor deviations attributed to optical aberrations and perspective effects. These laboratory demonstrations validate the DFI architecture as a highly capable and mature technology for future space-based FIR observatories.
\end{abstract}




\section{Introduction}

Far-infrared (FIR) radiation, typically defined as $25-350\,\mu\mathrm{m}$ ($1.20-0.85\,\mathrm{THz}$), carries roughly half of the light emitted by stars and accreting objects, because dust and gas enshrouding star formation and active galactic nuclei (AGN) reprocess shorter-wavelength emission to longer wavelengths \citep{blain:2002a, hocuk:2017a, tielens:2005a}. The band contains the dominant cooling lines of the interstellar medium (ISM), such as [CII], [OI], [OIII], and [NII], which trace gas from HII regions to the cold neutral medium \citep{tielens:1985a, ramospadilla:2023a}. High-ionization AGN tracers such as [NeV] and [OIV] are redshifted into the FIR at high redshift, enabling studies of the coevolution of AGN and star formation near the peak of cosmic star formation \citep{groves:2006a, spinoglio:2005a}. The FIR also hosts rotational transitions of CO, HD, and $\mathrm{H_2O}$, and water ice features, which probe molecular gas, total molecular mass, and the water content of protoplanetary disks \citep{linz:2021a}. In addition, the thermal emission of $\sim 10-100\,\mathrm{K}$ dust peaks in the FIR, making it the optimal band for observing the earliest stages of star and planet formation.

Despite this potential \citep{helmich:2009a, leisawitz:2023a}, the FIR remains one of the least explored bands. Atmospheric water vapor absorbs most incoming FIR radiation, requiring space-based observatories, while cryogenically cooled apertures and optics are required for background limited observations \citep{traub:1976a}. Detector sensitivity has historically been a further barrier, but TES \citep{irwin:2005a}, kinetic inductance \citep{mauskopf:2018a, zmuidzinas:2012a}, and quantum capacitance detectors \citep{echternach:2018a} now reach the noise levels needed for background-limited performance \citep{farrah:2019a}.


Additionally, FIR astronomy suffers from diffraction-limited spatial resolution two orders of magnitude worse than the surrounding near-infrared and submillimeter bands \citep{buinhas:2016a}, resulting in the so-called ``FIR gap.'' Using the Rayleigh criterion, the minimum angular separation between two point sources that can be distinguished using a circular aperture is \citep{born:1999a}:
\begin{equation}
    \label{eq:angres_circularap}
    \Delta \theta_{circ} \approx 1.22 \frac{\lambda}{D} \;,
\end{equation}
where $\lambda$ is the wavelength of light and $D$ is the diameter of the aperture. The longer FIR wavelengths demand proportionately larger primary mirrors; for example, achieving an angular resolution of 0.1 arcsec at $100 \,\mathrm{\mu m}$ would require a primary mirror $\sim 250 \,\mathrm{m}$ in diameter.

A promising alternative is a spatial interferometer consisting of two smaller apertures that collect and send light to a common beam combiner. The diffraction-limited angular resolution of a spatial interferometer is \citep{born:1999a}:
\begin{equation}
    \label{eq:angres_interferometer}
    \Delta \theta_{int\!f} = \frac{\lambda}{2 B} \;,
\end{equation}
where $B$ is the baseline separation between apertures. With an interferometer, the required resolution can be obtained with two smaller telescopes separated by $\sim 100 \,\mathrm{m}$. While spatial interferometry is a mature technique, it has yet to be performed in space. However, given ongoing feasibility studies, mission concepts, and planned missions, a FIR spatial interferometer is a natural progression for the field.

Much of the merit of FIR astronomy is due to its rich complement of spectral lines and continuum emission features. Consequently, an ideal instrument would combine a spatial interferometer with a broadband spectrometer. The Fourier transform spectrometer (FTS; \cite{davis:2001a}) is a common choice for FIR applications (e.g., the Herschel Space Telescope \cite{pilbratt:2010a}) for its high throughput, multiplexing advantage, and well-characterized line shape. Combining a spatial interferometer and an FTS produces a double-Fourier interferometer (DFI) that allows spectral imaging over a broad bandwidth with enhanced spatial resolution. Such an instrument has been considered in multiple mission concepts (e.g., Far-infrared interferometer \citep{helmich:2009a}, the Submillimeter Probe of the Evolution of Cosmic Structure \citep{leisawitz:1999a}, and the Space Infrared Interferometric Telescope \citep{leisawitz:2007a}) and studied via laboratory testbeds (e.g., \cite{itoh:1986a, ohta:2007a, leisawitz:2003a, grainger:2012a}) and balloon-borne prototypes (e.g., \cite{rinehart:2014a, kato:2010a}). Despite these efforts, there is no published work demonstrating the full capabilities of a DFI. Recently, a testbed developed by the Astronomical Instrumentation Group at the University of Lethbridge has been used to demonstrate key DFI capabilities independently, including aperture synthesis on a single pixel, sensitivity to spatially dependent spectral variations, and wide-field imaging.

In this paper, we describe the DFI laboratory testbed in \S~\ref{sec:Instrument_Summary} and derive both the ideal instrument response and non-ideal instrumental effects in \S~\ref{sec:instrument_response}. We use these derivations to characterize data acquisition and calibration procedures in \S~\ref{sec:processing}. In \S~\ref{sec:results} we present results that demonstrate the key observing capabilities of the DFI including aperture synthesis for one and two-dimensional targets, sensitivity to spatially dependent spectral variations of the source, and wide-field imaging using multiple detector pixels. We discuss these results and provide explanations for the slight disagreement between experimental results and theoretical predictions. Finally, we provide concluding remarks in \S~\ref{sec:conclusion}. 

\section{Instrument Summary}
\label{sec:Instrument_Summary}

\begin{table}
    \centering
    \begin{threeparttable}
    \caption{Parameters for the interferometer system (adapted from \cite{scott:2023a}).}
    \label{tab:instrument_parameters}
        \begin{tabular}{ l c r }
        \hline \hline
            Parameter & Value & Units  \\
            \hline
            
            collimating mirror focal length                    &      $1,200$ &  mm  \\
            field of view (pixel/array)\tnote{\dag}            &     $14/75$ &  mm  \\
            field of view (pixel/array)                        &     $40/215$ &  arcmin  \\
            magnification                                      &       $-0.2$ &  -   \\
    
            detector feed horn f/\#                                     &        $4.5$ &  -   \\
            optical NEP                                        &        $2.8$ &  pW/$\sqrt{\mathrm{Hz}}$ \\
            detector time constant                             &        $0.93$ &  ms  \\
            sampling rate                                      &         $13$ &  kHz \\
            bandwidth                                          &  $0.3 - 2.1$ &  THz \\
    
            baselines                                          &  $135 - 310$ &  mm  \\
            spectral resolution\tnote{\ddag}                   &       $0.19$ &  GHz \\
            spatial resolution ($\nu = 1.2\,\mathrm{THz}$)\tnote{\dag}  &        $1.0$ &  mm  \\
            spatial resolution ($\nu = 1.2\,\mathrm{THz}$)     &        $3.0$ &  arcmin  \\
            
            \hline \hline
        \end{tabular}
    \begin{tablenotes}
        \item[\dag] Defined on the source plane.
        \item[\ddag] Maximum spectral resolution for an on-axis point-source.
    \end{tablenotes}
    \end{threeparttable}
\end{table}

Instrument parameters and performance metrics are summarized in Table~\ref{tab:instrument_parameters} which are adapted from a dedicated instrument performance paper \citep{scott:2023a}. A summary of the optical layout, observing process, and hardware configuration follows below. 

\begin{figure}[htbp]
    \centering
    \begin{subfigure}[b]{0.95\textwidth}
        \includegraphics[width=\textwidth]{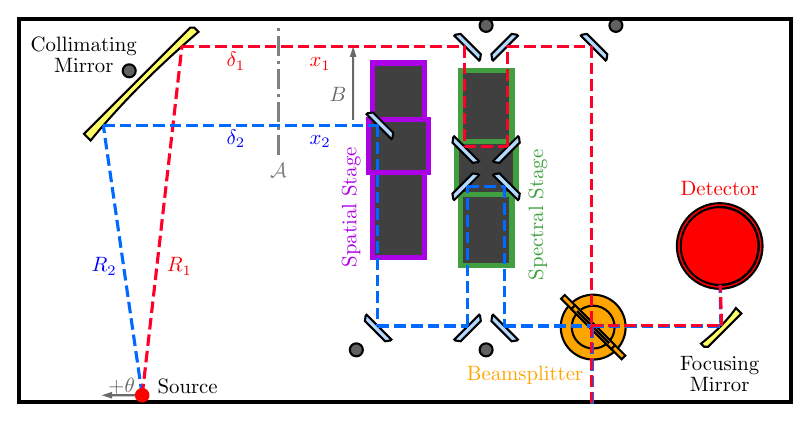} 
        \caption{DFI configuration used for this work. Input light is sampled by two apertures conceptually located on a synthetic aperture plane $\mathcal{A}$ with centers separated by a baseline length $B$, then propagated along two paths with optical path lengths $x_1$ and $x_2$. The beam are combined using a second beamsplitter and imaged onto the detector plane. The baseline is controlled by the spatial stage, and OPD is controlled by the spectral stage. The resulting signal is an interference pattern recorded as a function of OPD at a particular baseline.}
        \label{fig:doubleFourier_diagram}
    \end{subfigure}
    \vfill
    \begin{subfigure}[b]{0.95\textwidth}
        \includegraphics[width=\textwidth]{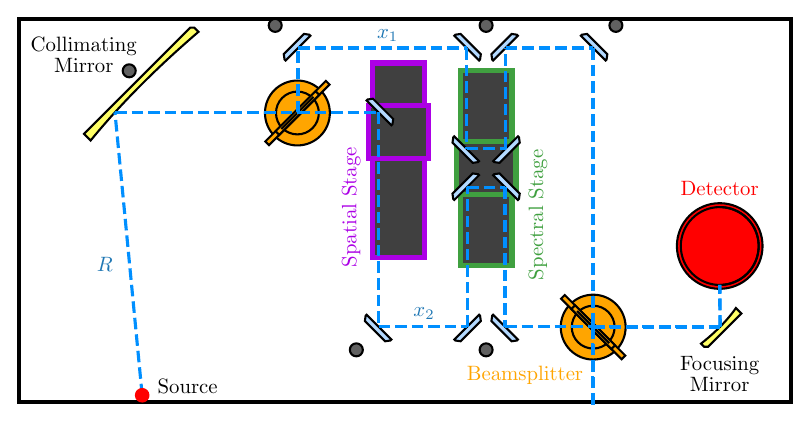} 
        \caption{FTS configuration used for this work. Input light is split by a 50:50 intensity beamsplitter then propagated along two paths with optical path lengths $x_1$ and $x_2$. The beams are combined using a second beamsplitter and imaged onto the detector plane. OPD is controlled by the spectral stage and the resulting signal is an interference pattern recorded as a function of OPD.}
        \label{fig:FTS_diagram}
    \end{subfigure}
    \caption{Diagrams of the testbed instrument showing to scale components on a $3\times1.5 \,\mathrm{m}$ optics table. The top panel shoes the instrument in DFI mode while the bottom panel shows the instrument in FTS mode.}
    \label{fig:testbed_diagrams}
\end{figure}

The testbed instrument architecture is illustrated in Fig.~\ref{fig:testbed_diagrams}. In the DFI mode (top panel), spherical waves emitted from point sources on the source plane are reflected by an off-axis parabolic (OAP) mirror to produce collimated plane waves, simulating a distant source. This light is sampled by two independent apertures consisting of circular flat mirrors oriented at $45^\circ$. These beams are relayed to a metal mesh beamsplitter \citep{ade:2006a}, for recombination and subsequently focused by a second OAP mirror onto the detector array.

We define an aperture plane positioned in front of the physical apertures, with a normal vector parallel to the intersecting rays from a point source on the optical axis. The interferometric baseline is defined as the separation between the centers of the two physical apertures when projected onto this plane. This baseline is adjusted using a high-precision Aerotech ALS20045 linear translation stage (the ``spatial stage''), providing a range from $135 \,\mathrm{mm}$ to $310 \,\mathrm{mm}$. The lower and upper limits are determined by the non-overlapping footprint of the apertures and the extent of the collimating mirror, respectively.

A double-rooftop mirror assembly, mounted on a second Aerotech stage (the ``spectral stage''), modulates the optical path difference (OPD) between the two beams. This configuration yields four units of OPD for every one unit of mechanical travel. The instrument can be converted to FTS mode by introducing an additional beamsplitter and mirror into the optical chain (Fig.~\ref{fig:testbed_diagrams}, bottom panel). In this configuration, FTS mode is functionally equivalent to DFI mode operating at zero baseline.

An observation sequence begins with the construction of a scene on the source plane, typically using a mask with geometric cutouts backlit by a thermal source at $1,000 \,^\circ\mathrm{C}$. Before interferometric data collection, photometric calibration is performed. The source is modulated by a mechanical chopper while one beam is blocked. The peak-to-peak amplitude of the modulated signal is averaged over multiple cycles to estimate the power reaching the detector through each aperture at every sampled baseline.

Interferometric measurements are initiated by setting the spatial stage to the minimum baseline and offsetting the spectral stage $2 \,\mathrm{mm}$ ($8 \,\mathrm{mm}$ OPD) from the Zero Path Difference (ZPD). The spectral stage then performs continuous scans at 6.7 mm/s over a $\pm 2$ mm range relative to ZPD. During travel, the stage emits a position-synchronized output (PSO) voltage signal every 0.015 mm that is recorded by the detector system, which simultaneously records incident intensity at $\sim 13,000 \,\mathrm{frames/s}$. This allows the system to record optical power as a function of OPD and baseline, capturing the temporal and spatial coherence of the source. This instrument response is derived in \S~\ref{subsec:Ideal_Instrument_Response}. A typical observation consists of $\ge 2,000$ forward and reverse scans per baseline. This process is repeated for all spatial positions and subsequently replicated in FTS mode.

The detector array consists of a $5 \times 5$ grid of feed horn coupled TES housed in a cryostat at an operating temperature of $\sim4 \,\mathrm{K}$ \citep{benson:2025a}. The 25 sensors are distributed across four channels (A–D) and read out via time-domain multiplexing with each sensor assigned a number (0-6) denoting its position in the readout sequence. Unique to this system, each sensor is maintained at its critical temperature through a PID-controlled heater wire. While this feedback loop results in a relatively high Noise Equivalent Power (NEP) of $\sim2.8 \,\mathrm{pW/\sqrt{Hz}}$, it prevents detector saturation from laboratory background radiation and allows for precise parameter tuning. Additionally, the detector assembly includes an external channel to record high/low voltage signals, facilitating synchronization between the detector data and external metrology.

The optical design was optimized for the central sensor (A0), achieving a $14 \times 14 \,\mathrm{mm}$ ($40 \times 40 \,\mathrm{arcmin}$) field of view (FOV) on the source plane to maximize throughput. Array characterization confirms that A0 and adjacent sensors (notably A3) maintain high source plane coupling efficiency across all aperture positions. 

The instrument operates across an optical bandwidth of $0.3 - 2.1 \,\mathrm{THz}$. The low-frequency cutoff is established by the metal mesh beamsplitter transmission profile and the single-mode cutoff of the detector feed horns. Higher frequencies undergo gradual attenuation due to the degrading coupling efficiency of the detectors at shorter wavelengths. Measured blackbody spectra are further characterized by atmospheric water vapor absorption features with significant attenuation in the neighborhood of $1.2 \,\mathrm{THz}$ and $1.7 \,\mathrm{THz}$.

\section{Instrument Response}
\label{sec:instrument_response}

\subsection{Ideal Instrument Response}
\label{subsec:Ideal_Instrument_Response}

\subsubsection*{Non-zero Baseline}

The electric fields incident on the detector resulting from a point-source on the source plane at position $\vec{\theta}$ with wavenumber $k = 1/\lambda = \nu/c$ are:
\begin{equation}
    \label{eq:electric_fields}
    \begin{split}
        E_1(\vec{\theta}, k) & = E_{\mathcal{A}1}(\vec{\theta}, k) e^{ i 2\pi k (x_1 + \vec{B}\cdot\vec{\theta} - ct) } \\
        E_2(\vec{\theta}, k) & = E_{\mathcal{A}2}(\vec{\theta}, k) e^{ i 2\pi k (x_2 - ct) }
    \end{split}
\end{equation}
where the subscript indicates which arm of the interferometer the wave propagates through. Each wave travels an optical path length $x_i$ from the aperture plane $\mathcal{A}$ to the detector. The position of the aperture plane is chosen for convenience such that the distance from the focal point of the collimating mirror to the aperture plane is an integer multiple of the wavelength of light being considered (i.e., $R_i + \delta_i = n\lambda$ in Fig.~\ref{fig:doubleFourier_diagram}), simplifying the phase terms in equation~\ref{eq:electric_fields}. The electric fields enter the interferometer with amplitudes $E_{\mathcal{A}i}$ through apertures separated by the baseline vector $\vec{B}$. For a point source displaced from the optical axis by an angle $\vec{\theta}$, light entering $\mathcal{A}1$ is delayed with respect to $\mathcal{A}2$ by an optical path length $\vec{B}\cdot\vec{\theta}$. 

At the detector, the measured intensity is given by the time average of the sum of the electric fields through the two arms,
\begin{equation}
    \label{eq:DFI_response_point}
    \begin{split}
        I_{DFI}(\vec{B}, z; \vec{\theta}, k) & = \left\langle \left| E_1(\vec{\theta}, k) + E_2(\vec{\theta}, k) \right|^2 \right\rangle \\
        & = \left\langle \left| E_1(\vec{\theta}, k) \right|^2 \right\rangle + \left\langle \left| E_2(\vec{\theta}, k) \right|^2 \right\rangle + 2\left\langle \left| E_1^*(\vec{\theta}, k) E_2(\vec{\theta}, k) \right| \right\rangle \\
        & = I_1(\vec{\theta}, k) + I_2(\vec{\theta}, k) + 2\sqrt{I_1(\vec{\theta}, k) I_2(\vec{\theta}, k)} \; e^{ i 2\pi k (z - \vec{B}\cdot\vec{\theta}) }
    \end{split}
\end{equation}
where $z = x_2 - x_1$ is the instrumental optical path difference (OPD) and $I_i(\vec{\theta}, k)$ are the intensities at the detector through each arm of the interferometer. 

Under ideal conditions, the intensities are balanced (i.e., $I_1(\vec{\theta}, k) = I_2(\vec{\theta}, k) = I(\vec{\theta}, k)$). Additionally, the bias in the signal is typically subtracted leaving only the term that varies with OPD. With these considerations, and integrating over wavenumber and source plane $\Omega$, equation~\ref{eq:DFI_response_point} becomes
\begin{equation}
    \label{eq:DFI_response_full}
    I_{DFI}(\vec{B}, z) = \int_\Omega \int_{-\infty}^{\infty} I(\vec{\theta}, k) \; e^{ i 2\pi k z } \; e^{ -i 2\pi k \vec{B}\cdot\vec{\theta} } \; dk \; d\vec{\theta} \;,
\end{equation}
where we have defined the source spectrum to be real and symmetric (i.e., $I(\vec{\theta}, k) = I(\vec{\theta}, -k)$). Equation~\ref{eq:DFI_response_full} is functionally equivalent to the mutual coherence function \citep{wolf:1955a}, a generalization of the van Cittert-Zernike theorem extended to electromagnetic fields of arbitrary spectral width \citep{wolf:1955a}. The instrument response is thus a two dimensional Fourier transform over the source intensity distribution at a given wavelength (probing spatial coherence), and a one dimensional inverse Fourier transform over OPD (probing temporal coherence).

We define the complex visibility as
\begin{equation}
    \label{eq:visibility_complex}
    \begin{split}
            \gamma(\vec{B}, k) & = \frac{ \int_\Omega I(\vec{\theta}, k) \; e^{ -i 2\pi k \vec{B}\cdot\vec{\theta} } \; d\vec{\theta} }{ I(k) } \\
            & = \left| \gamma(\vec{B}, k) \right| e^{i \phi_\gamma(\vec{B}, k)} \;,
    \end{split}
\end{equation}
where 
\begin{equation}
    \label{eq:integrated_source_spectrum}
    I(k) = \int_\Omega I(\vec{\theta}, k) \; d\vec{\theta}
\end{equation}
is the integrated source spectrum. This normalization ensures $ 0 \le \left| \gamma(\vec{B}, k) \right| \le 1$ and effectively removes spectral amplitude variations from the source. Rewriting equation~\ref{eq:DFI_response_full} using equation~\ref{eq:visibility_complex}, we obtain
\begin{equation}
    \label{eq:DFI_response_visibility}
    I_{DFI}(\vec{B}, z) = \int_{-\infty}^{\infty} I(k) \left| \gamma(\vec{B}, k) \right| \; e^{ i \left( 2\pi k z +  \phi_\gamma(\vec{B}, k) \right) } \; dk \;.
\end{equation}
This form of the DFI response can be used to obtain some intuition about how the source intensity distribution is encoded in the response. 

For a single monochromatic component $k = k_0$, equation~\ref{eq:DFI_response_visibility} becomes
\begin{equation}
    \label{eq:DFI_response_mono}
    I_{DFI}(\vec{B}, z; k_0) = 2I(k_0) \, \left| \gamma(\vec{B}, k_0) \right| \cos \left( 2\pi k_0 z - \phi_{\gamma}( \vec{B}, k_0 ) \right) \;,
\end{equation}
which represents a cosine interference pattern as a function of OPD with an amplitude proportional to the source spectral amplitude at $k_0$. This pattern is further modulated by the amplitude of the complex visibility $\left| \gamma(\vec{B}, k_0) \right|$ and is offset along the OPD axis by the complex visibility phase $\phi_{\gamma}( \vec{B}, k_0 )$. If the source is a point source, the complex visibility amplitude is unity and the pattern is shifted by $\Delta z = \vec{B} \cdot \vec{\theta}$ along the OPD axis. For an extended monochromatic source, where the response is the sum of many point sources, the complex visibility amplitude tends to decrease and the pattern tends to shift according to the center of mass. The full response is the sum of many spectral components. 

\subsubsection*{Zero Baseline}

Setting $\vec{B}=0$ and integrating over the source-plane, equation~\ref{eq:DFI_response_full} becomes
\begin{equation}
    \label{eq:FTS_response}
    I_{FTS}(z) = I_{DFI}(0, z) = \int_{-\infty}^{\infty} I(k) \; e^{ i 2\pi k z } \; dk \;, 
\end{equation}
which reproduces the response of an FTS. In practice, FTS mode cannot be achieved by overlapping apertures since one aperture will physically block the other. We overcome this limitation by introducing another mirror and beamsplitter into the optical chain as shown in Fig.~\ref{fig:FTS_diagram}. As a consequence of using only one aperture, throughput of the instrument in FTS mode is reduced to half that of DFI mode. 

\subsection{Instrumental Effects}

In practice, the ideal response is modified by various instrumental effects. To ensure accurate source reconstruction, we must account for intensity imbalances between the interferometer arms, phase-center offsets, and frequency-dependent phase errors.

\subsubsection*{Intensity Imbalance}

The intensities arriving at the detector from both arms of the interferometer are not equal (i.e., $I_1(\vec{\theta}, k) \ne I_2(\vec{\theta}, k)$). For the testbed instrument, the electric field at the aperture plane is related to the distance between the source plane and collimating mirror $R_i$ (i.e., $E_{\mathcal{A}i}(\vec{\theta}, k) \propto E_{src}(\vec{\theta}, k) / R_i$) and $R_1 \ne R_2$. A space-based instrument would not suffer from this effect since $R_i \gg R_2 - R_1$, however, asymmetries within the instrument may be present and require calibration.  

The intensity at the detector is modeled as 
\begin{equation}
\label{eq:photocal_intensity}
I_i = \int_{\Omega} \int_{-\infty}^{\infty} a_i \, \xi(\vec{\theta}, k) \, I(\vec{\theta}, k) \, dk \, d\Omega \;,
\end{equation}
where $\xi(\vec{\theta}, k)$ is the source plane sensitivity of the detector determined by the aperture and detector geometry, and $a_i$ is a scaling variable dependent on the aperture position. This model assumes the variations in intensity through each arm of the interferometer are spectrally independent and that the source plane sensitivity functions for each aperture are similar in shape. Photometric calibration measurements described in \S~\ref{sec:Instrument_Summary} are used to estimate $I_i$, and the instrument response is normalized by dividing by $\sqrt{I_1 I_2}$.

Limiting the FOV through $\xi(\vec{\theta}, k)$ imposes constraints on the $uv$~plane sampling required to avoid aliasing. If the instrument is sensitive to an angular diameter of $\theta_{fov}$, then sampling in the $uv$-plane should obey $\Delta u = \Delta v \le 1 / \theta_{fov}$. 

\subsubsection*{Phase Errors}

Instrumental phase errors can be decomposed into multiple components \citep{davis:2001a},
\begin{equation}
    \label{eq:instrumental_phase}
    \phi_{inst}(k) = \phi_{NL}(k) + \phi_L(k) + \phi_R(k) \;,
\end{equation}
which represent non-linear, linear, and random effects, respectively. The random term is due to effects including optics vibrations, air turbulence, and irregular response sampling which cannot be calibrated but may be reduced with averaging. The non-linear term is influenced primarily by the relative phase imbalance between the two beams upon transmission and reflection at the beamsplitter in addition to a potentially non-linear response of the detector. The linear component results from incorrectly sampling ZPD and residual linear phase error from the beamsplitter and detector.

Phase errors can be included in the instrument response model by multiplying $e^{i\phi_{inst}(k)}$ into equation~\ref{eq:DFI_response_full}. We note that these phase errors are expected to be independent of baseline as the dominant sources of error are the beamsplitter, detector, and OPD sampling.

\section{Data Processing}
\label{sec:processing}

A typical observation was described in \S~\ref{sec:Instrument_Summary}. The science data produced from these observations are time series of incident optical power as a function of OPD called interferograms. Individual interferograms are truncated to remove accelerations regions at their ends, ensuring the the data is sampled at uniform intervals of OPD. Interferograms are then averaged to improve the signal-to-noise ratio (SNR). Truncation and alignment of interferograms is facilitated using the PSO signal. For the data used in this work, truncating the interferogram results in an experimental spectral resolution of $\sim 25 \,\mathrm{GHz}$, though the instrument can achieve $0.19\,\mathrm{GHz}$ resolution at the cost of significantly longer observation time. An example of an averaged interferogram after bias subtraction is shown in Fig.~\ref{fig:correctedintfg}. 

\subsection*{Non-linear Phase Correction}

\begin{figure}
	\centering
	\includegraphics[width=0.8\linewidth]{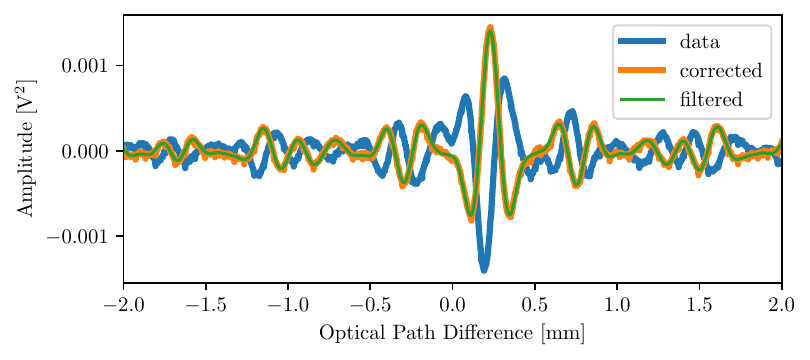}
	\caption[Phase Corrected Interferogram]{Phase corrected interferogram. The blue curve is an averaged interferogram obtained when observing a $1\,\mathrm{mm}$ , approximately on-axis, slit backlit by a $1,100\,\mathrm{^\circ C}$ black body, using a $135\,\mathrm{mm}$ baseline. The orange curve shows the same interferogram after phase correction, which results in the green curve after passing through a $300 - 2,200\,\mathrm{GHz}$ optical frequency band-pass filter. Note the high degree of symmetry after correcting for non-linear phase errors.}
	\label{fig:correctedintfg}
\end{figure}

\begin{figure}[t]
	\centering
	\includegraphics[width=0.8\linewidth]{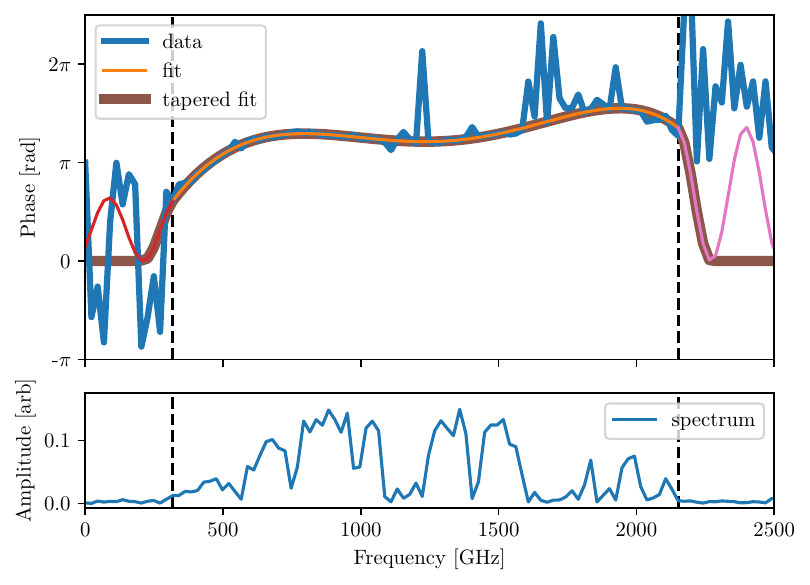}
	\caption[Polynomial Fit to Non-linear Phase]{Polynomial fit to non-linear phase. The blue curves in the top and bottom panels represent the calculated phase and amplitude spectra of the data from Fig.~\ref{fig:correctedintfg}, respectively. The phase data residing within the instrument bandwidth (vertical dashed lines), is fit to a fifth degree polynomial, which provides a good result without over fitting the noise. Each data point in the phase spectrum is weighted by its corresponding amplitude in the amplitude spectrum to ensure high signal regions influence the fit. Note the fit matches the data quite well in regions with high amplitude. The fit is tapered to zero outside of the instrument bandwidth using sinusoids (red and pink curves), that satisfy boundary conditions such that their value and slope are equal to the polynomial's value and slope at the point where they intersect at $300\,\mathrm{GHz}$ and $2,200\,\mathrm{GHz}$.}
	\label{fig:phasefit}
\end{figure}

Since the ideal instrument response is a sum of cosine functions, when observing a point source (or a slightly extended symmetric source), the resulting interferogram should be symmetric about some point on the OPD axis. Deviations from symmetry are due to phase errors. We use this fact to estimate a phase correction product that can be applied to all interferograms recorded in DFI mode. A $1\,\mathrm{mm}$ target (see Fig.~\ref{fig:1mm_slit}) is placed at the center of the source plane and an interferogram is obtained using the smallest ($135 \,\mathrm{mm}$) baseline configuration. The combination of a small source observed with the smallest baseline ensures high visibility fringes, while placing it near the center of the source plane ensures maximum detector sensitivity. This strategy maximizes SNR when obtaining the phase calibration product. The resulting interferogram is the blue curve in Fig.~\ref{fig:correctedintfg}. 

After Fourier transformation with respect to OPD, the result is a complex spectrum with the form
\begin{equation}
    \label{eq:complex_spectrum}
    \mathcal{F}_z \left\{ I_{DFI}(\vec{B}, z) \right\}  = I(k) \left| \gamma(\vec{B}, k) \right| \; e^{ i  \phi_\gamma(\vec{B}, k) } \, e^{i \phi_{inst}(k) } \;,
\end{equation}
obtained from equation~\ref{eq:DFI_response_visibility} in connection with equation~\ref{eq:instrumental_phase}. Since the target is small and observed at the shortest baseline $\left| \gamma(\vec{B}, k) \right| \approx 1$, and since the target is symmetric $\phi_\gamma(\vec{B}, k) = 2\pi k\vec{B}\cdot\vec{\theta}$. That is, the complex visibility phase is a linear function of wavenumber and the non-linear components of the observed phase are due to instrumental effects. 

The observed phase spectrum, computed as the $\phi_{obs}(k) = \arctan(Im(k)/Re(k))$ of the real and imaginary components of the Fourier transform, is shown as the blue curve in the top panel of Fig.~\ref{fig:phasefit}. The lower panel shows the amplitude spectrum of the Fourier transform which represents $I(k) \left| \gamma(\vec{B}, k) \right|$, the visibility modulated source spectrum. In this case, the visibility is close to unity over the full bandwidth and the shape of the spectrum is strongly affected by instrumental and atmospheric effects \citep{scott:2024a}. 

Prominent absorption features across the band are accurately captured by an atmospheric model (900\,mbar, 21.5\,$^\circ\mathrm{C}$, and 14\% relative humidity over a 5.4\,m path), consistent with laboratory measurements. Additional losses occur at frequencies $\ge 1.5\,$THz due to the beamsplitter and cryostat entrance membrane, while the jagged appearance away from these features are consistent with the $n\lambda^2$ multimodal throughput of the rectangular detector waveguides. Higher-frequency attenuation is driven by two main components: a detector response time that acts as a $\sim 2\,$THz low-pass optical filter, and a more dominant drop in relative spectral response that decreases approximately linearly down to nearly zero at $\sim 1.6\,$THz, attributed to the geometry of the detector integrating cavity.

The observed phase is fit using a fifth degree polynomial which conforms well to the data without fitting the noise. Samples are weighted by their corresponding spectral amplitude so that high SNR points contribute more to the fit. In performing the Fourier transform, one can select which point of the interferogram is considered ZPD which affects the resulting phase spectrum. ZPD was selected in order to minimize the linear term in the fitted polynomial, intentionally ignoring any linear phase effects. The fitted polynomial is then tapered to zero at both ends of the instrument band by cosine functions subject to the boundary conditions that their value and slope must match the polynomial. The final phase correction product $\phi_{fit}(k) \approx \phi_{NL}(k)$, is represented by the brown curve in Fig.~\ref{fig:phasefit}.

Multiplying the complex spectrum (equation~\ref{eq:complex_spectrum}) by $e^{-i\phi_{fit}(k)}$ and inverse Fourier transforming back into OPD space produces a symmetrical interferogram, as shown by the orange curve in Fig.~\ref{fig:correctedintfg}. This phase correction product was further validated by applying it to interferograms obtained from other baselines and using other symmetric targets, all of which produced highly symmetric interferograms, as required. 

\subsection*{Phase-center Calibration}

\begin{figure}[t]
	\centering
	\includegraphics[width=0.8\linewidth]{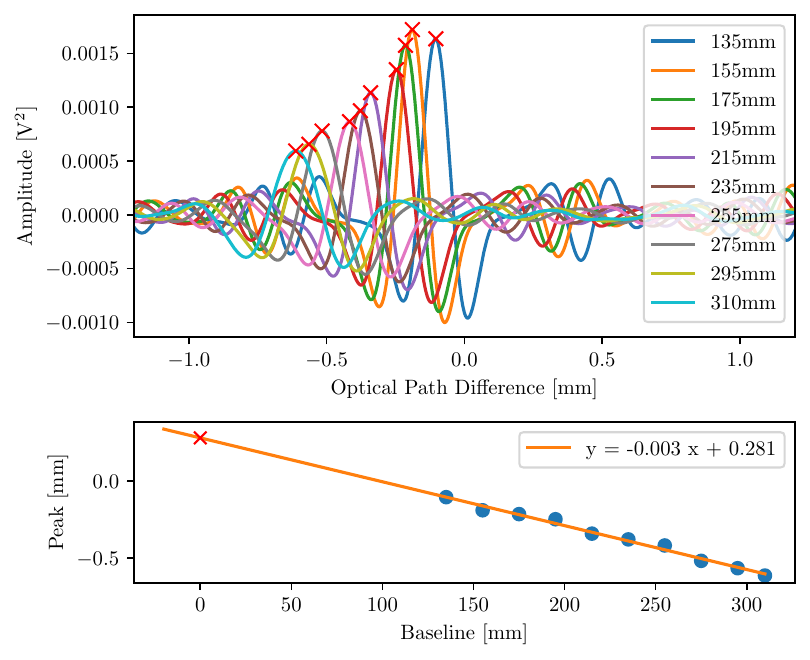}
	\caption[Phase Center Calibration]{Phase center calibration. The top panel shows the phase corrected DFI response obtained at multiple baselines when viewing a $1\,\mathrm{mm}$ source. Peaks of the sinc-like functions are identified and plotted as a function of baseline in the lower panel. These data are fit using a linear equation producing a y-intercept of $0.28 \pm 0.03\,\mathrm{mm}$. This value is identified as an uncorrected OPD offset in connection with Eq.~\ref{eq:phase_center_offset}, and is subtracted from the OPD axis as part of the calibration procedure.}
	\label{fig:opdoffsetcal}
\end{figure}

The remaining systematic phase error is a component linearly dependent on wavenumber (i.e., $\phi_L(k) = 2\pi k \delta z$). Also contributing to the measured phase is a linear term associated with the source position (i.e., $\phi_\gamma(\vec{B}, k) = 2\pi k \vec{B} \cdot \vec{\theta}$). Due to the phase-shift property of Fourier transforms, a linear phase shift in the frequency domain produces a constant offset in the OPD domain, 
\begin{equation}
    \label{eq:phase_center_offset}
    z_{offset} = \vec{B}\cdot\vec{\theta} + \delta z \; .
\end{equation}
By inferring the offset when $\vec{B}=0$, we can estimate residual offsets due to ZPD sampling error and any uncorrected linear phase error.

In practice, the baseline cannot be set to zero in DFI mode. However, the peak of an interferogram after non-linear phase correction can be estimated over a range of baselines. By fitting the peak positions as a function of baseline using a linear function in connection with equation~\ref{eq:phase_center_offset}, the y-intercept can be estimated and identified as a systematic OPD offset $\delta z$. By subtracting $\delta z$ from the nominally constructed OPD axis, the origin of the corrected axis corresponds to the phase center of the instrument (i.e., ZPD of a point source on the optical axis). After applying this correction, the data can be modeled using equation~\ref{eq:complex_spectrum} with the instrumental phase error dominated by random phase error which ideally approaches zero with sufficient averaging (i.e., $\phi_{inst}(k) \approx \phi_R(k) \rightarrow0 $).

\subsection*{$uv$-Plane Projection}

Following phase and photometric calibration, the data has the form,
\begin{equation}
\label{eq:complex_spectrum_cal}
    \begin{split}
        \mathcal{F}_{z,cal} \left\{ I_{DFI}(\vec{B}, z) \right\} & = I_{n}(k) \left| \gamma(\vec{B}, k) \right| \; e^{ i  \phi_\gamma(\vec{B}, k) } \\
        & = I_{n}(k) \gamma(\vec{B}, k) \;.
    \end{split}
\end{equation}
That is, the data are now the product of the complex visibility encoding the source intensity distribution (see equation~\ref{eq:visibility_complex}) and the normalized source spectrum. The normalized spectrum results from the photometric calibration step, and in connection with equation~\ref{eq:photocal_intensity}, is given by $I_n(k) = I(k) \, I_i/a_i$. Note that this data can be represented as ordered pairs for real and imaginary values.  

In principle, one can now construct a $uv$-plane for each spectral component $k$, and project the complex values from equation~\ref{eq:complex_spectrum_cal} into this space using the mapping
\begin{equation}
    \label{eq:uv_projection}
    \langle u, v \rangle = k \langle B_x , B_y \rangle = k \left| \vec{B} \right| \langle \cos(\theta_{\vec{B}}) , \sin(\theta_{\vec{B}}) \rangle \;,
\end{equation}
where $\theta_{\vec{B}}$ is the rotation angle between the baseline vector and some reference axis. Typically, only ten baseline lengths are measured for each baseline orientation, resulting in sparse $uv$ sampling. When the source intensity distribution is spectrally uniform (i.e., has the same shape at all wavelengths), the complex visibility will be the same at all wavelengths. However, the complex visibility is scaled by the normalized source spectrum and values from different spectral components cannot be immediately combined to obtain denser sampling.

After phase and photometric calibration of the FTS data (see equation~\ref{eq:FTS_response}), we obtain
\begin{equation}
\label{eq:FTS_spectrum_cal}
        \mathcal{F}_{z,cal} \left\{ I_{FTS}(z) \right\} = I_{n}(k) \;,
\end{equation}
which can be used to isolate the complex visibility from equation~\ref{eq:complex_spectrum_cal}. Now, the $uv$-plane can be super sampled by combining the complex visibility measurements from all spectral components and baselines.

\subsection*{Image Recovery}

The densely sampled $uv$-plane can now be inverse Fourier transformed to recover the source intensity distribution (see equation~\ref{eq:visibility_complex}),
\begin{equation}
    \label{eq:image_synthesis}
    \mathcal{F}^{-1}_{uv}   \left\{ \gamma(\vec{B}, k) \right\} =\frac{ I(\vec{\theta}, k) }{ I(k) } \;,
\end{equation}
which gives the normalized source intensity distribution. Unlike the uniform sampling in OPD, which allows for the efficient use of the fast Fourier transform (FFT; \cite{cooley:1965a}) algorithm when computing equation~\ref{eq:complex_spectrum_cal}, the $uv$-plane is generally sampled irregularly. As such, equation~\ref{eq:image_synthesis} must be computed using the less efficient direct Fourier transform (DFT; \cite{perley:1989a}). Alternatively, $uv$ samples can be projected onto uniform sampling intervals using regridding methods commonly used in radio astronomy aperture synthesis. In our work, we employ the convolution regridding method \citep{perley:1989a, scott:2024a}. Regridding allows for the use of the FFT, and this approach is generally faster than DFT alone when the number of $uv$ samples is sufficiently high, as would be expected for real astronomy applications. 

The image obtained from equation~\ref{eq:image_synthesis} is called the ``dirty image.'' It is modeled as the convolution of the true source intensity distribution with the instrumental point spread function (PSF) where the PSF is obtained by computing equation~\ref{eq:image_synthesis} with the complex visibility values set to unity at the sampled spatial frequencies and zero everywhere else. The dirty image is characterized by distortions resulting from missing spatial frequencies in the image reconstruction, typically resulting in excess wave features in the reconstructed image. Often the dirty image is post processed using PSF deconvolution algorithms such as CLEAN \citep{clark:1980a} or maximum entropy methods \citep{gull:1984a}. Such algorithms were not used in this work since the target scenes are simple and comparisons of experimental and theoretical dirty images provide compelling verification of the DFI technique on their own. However, the application of the CLEAN algorithm to these results can be found in \cite{scott:2024a}.

As a final processing step, the dirty image is divided by the source plane sensitivity function $\xi(\vec{\theta},k)$ which removes image variations that result from the instrumental coupling efficiency to different points in the source plane. Note that $\xi(\vec{\theta},k)$ may change with baseline vector, in which case a synthetic $\xi(\vec{\theta},k)$ must be constructed from the average source plane sensitivity over the sampled baselines. 

\section{Experimental Results and Analysis}
\label{sec:results}

\begin{figure}
     \centering
     \begin{subfigure}[b]{0.3\textwidth}
         \centering
         \includegraphics[width=\textwidth]{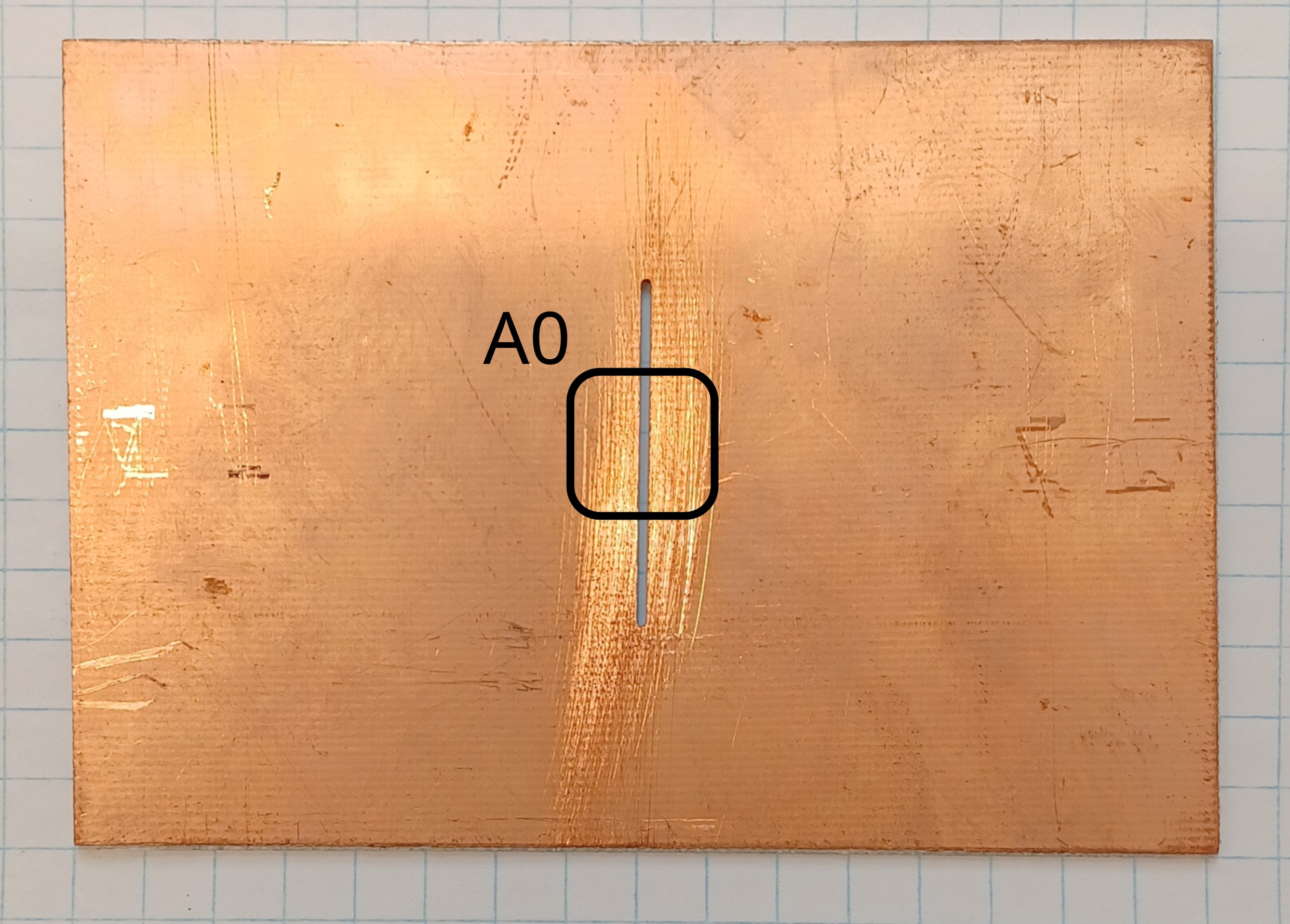}
         \caption{$1\,\mathrm{mm}$ slit. \vspace{0pt}}
         \label{fig:1mm_slit}
     \end{subfigure}
     \begin{subfigure}[b]{0.3\textwidth}
         \centering
         \includegraphics[width=\textwidth]{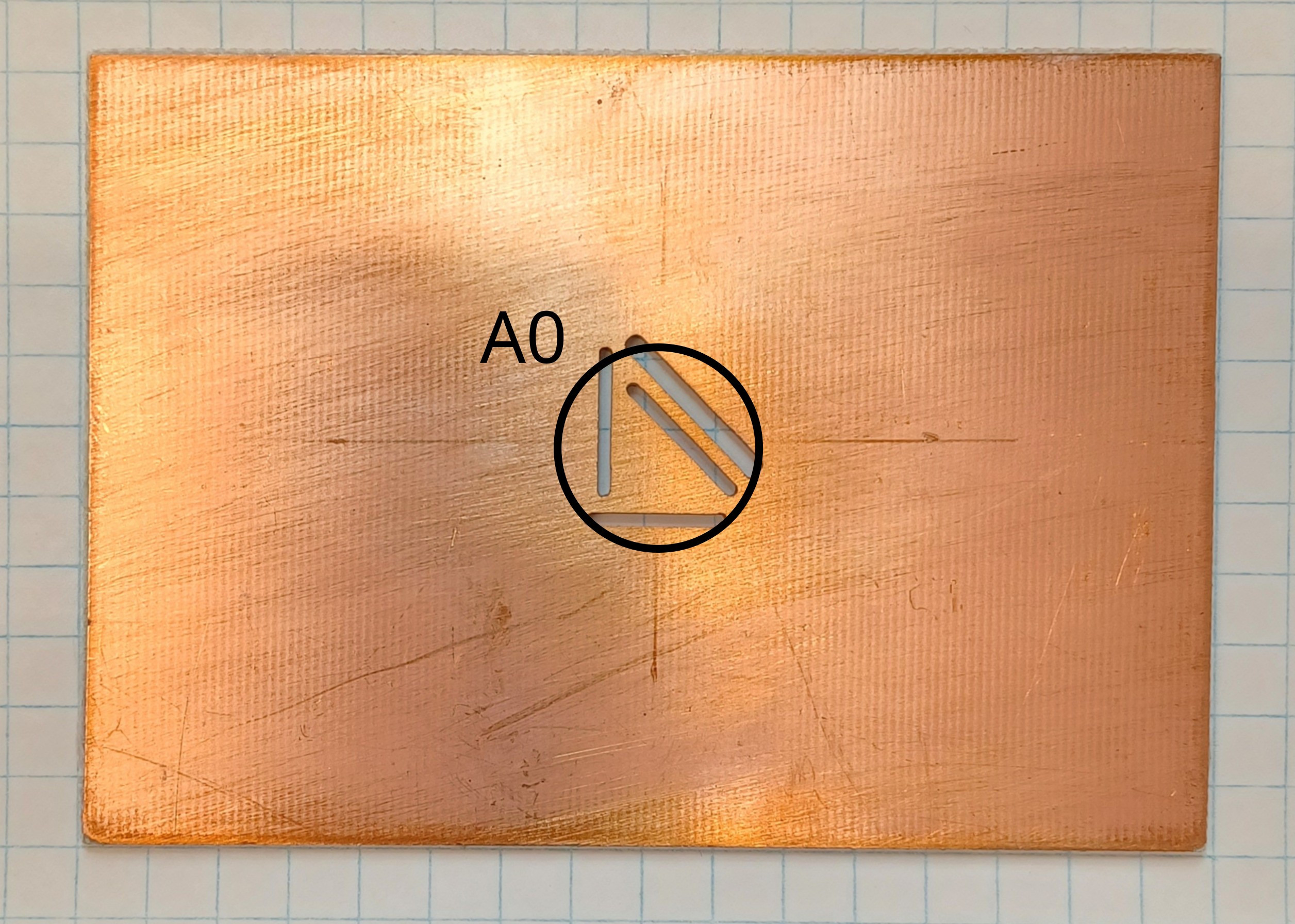}
         \caption{2D mask of $1\,\mathrm{mm}$ and $2\,\mathrm{mm}$ slits.}
         \label{fig:2d_mask}
     \end{subfigure}

     \begin{subfigure}[b]{0.3\textwidth}
         \centering
         \includegraphics[width=\textwidth]{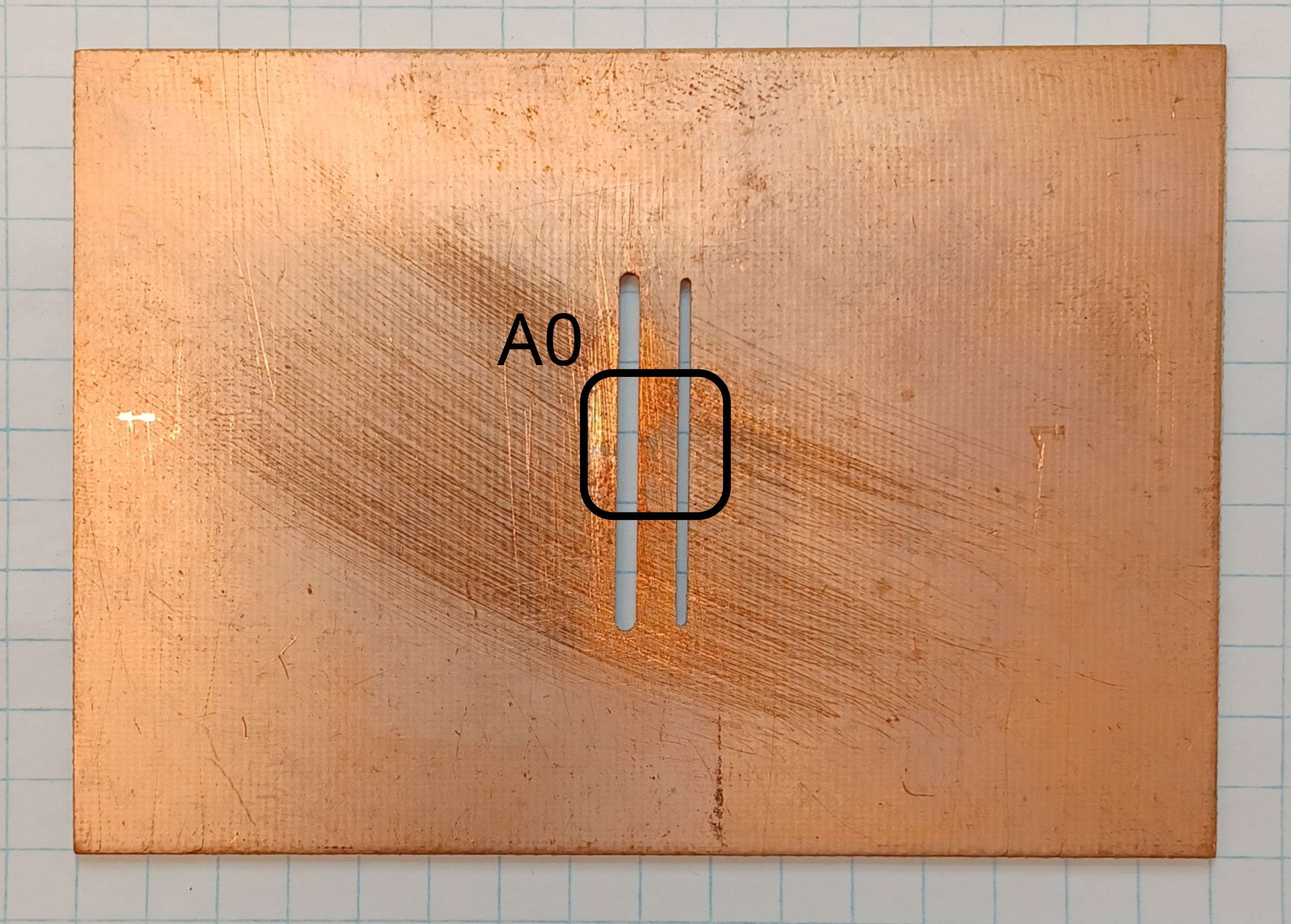}
         \caption{$2\,\mathrm{mm}$ and $1\,\mathrm{mm}$ slit separated by $5\,\mathrm{mm}$.}
         \label{fig:2slit_5mm}
     \end{subfigure}
     \begin{subfigure}[b]{0.3\textwidth}
         \centering
         \includegraphics[width=\textwidth]{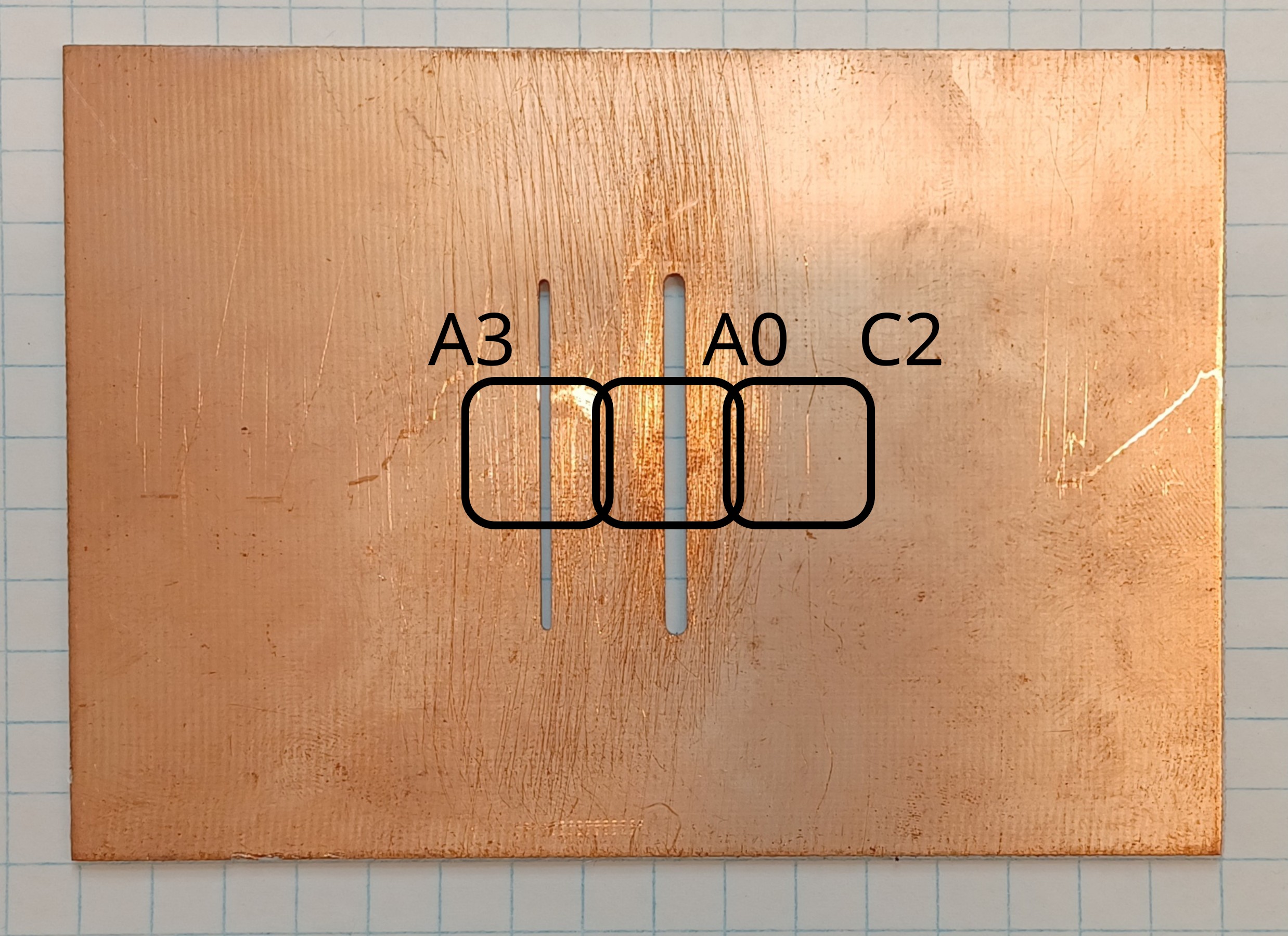}
         \caption{$1\,\mathrm{mm}$ and $2\,\mathrm{mm}$ slit separated by $12\,\mathrm{mm}$.}
         \label{fig:2slit_12mm}
     \end{subfigure}
        \caption{Source masks used to demonstrate the DFI observation technique. Annotations show the approximate projected FOV of detectors, which is circular in \ref{fig:2d_mask} due to mask rotation.}
        \label{fig:aperture_masks}
\end{figure}

\begin{figure}
	\centering
	\includegraphics[width=0.4\textwidth]{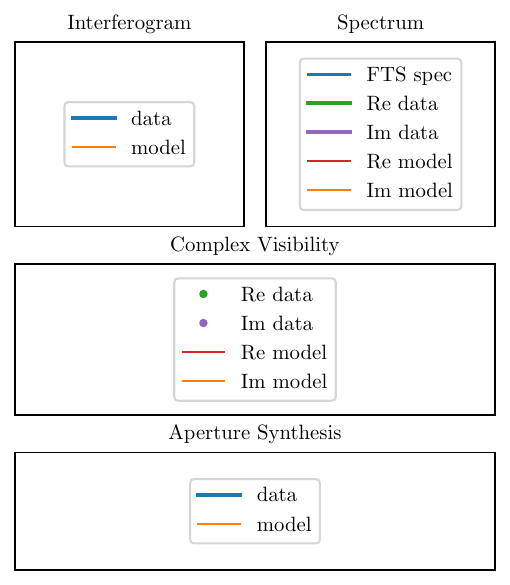}
	\caption[Multi-panel Figure Legend]{Multi-panel figure legend. The color scheme used in the results figures presented throughout this work are consistent with those indicated in this figure. Model curves denote the forward-simulated instrument response derived from the assumed physical source geometry, rather than a fitted parameter model.}
	\label{fig:legend}
\end{figure}

Scenes were constructed by placing masks on the source plane which were backlit using a thermal source (see \S~\ref{sec:Instrument_Summary}). The masks used in this work are shown in Fig.~\ref{fig:aperture_masks}. Results obtained for all the various observations will be presented in a similar fashion using four subfigures. To improve readability, the legends for these subfigures are presented in Fig.~\ref{fig:legend}. The top-left panel shows calibrated measured interferograms and theoretical interferograms for a subset of baselines. The top-right panel shows the resulting complex spectra for the corresponding measured and simulated interferograms. Also included is the measured FTS (zero baseline) spectrum which is used as the source spectrum for simulation purposes. The middle panel shows measured and simulated complex visibilities. The bottom panels shows the experimental and simulated dirty image in addition to the source geometry used for simulation purposes.  

In each case, the source panel displays the input physical source geometry (adjusted for perspective effects but not optimized with fitting tools). We perform forward modeling using this source and the measured FTS spectrum to generate theoretical interferograms, complex spectra, visibilities, and dirty images.

\subsection{One-Dimensional Sources}
\label{subsec:1d_sources}

The source masks shown in Figs.~\ref{fig:1mm_slit},~\ref{fig:2slit_5mm}, and \ref{fig:2slit_12mm} have essentially no spatial variation along the vertical direction within the detector FOV and can be regarded as one dimensional targets with spatial variation along the horizontal direction. This has two advantages: first, the relevant information can be measured with a single baseline orientation, and second, flux from the target along the vertical direction can be added effectively increasing the amplitude of the signal. These conditions allow for efficient observations while still demonstrating the enhanced spatial resolution of the DFI. 

\subsubsection*{Enhanced Spatial Resolution}

\begin{figure}
	\centering
	\includegraphics[width=0.8\linewidth]{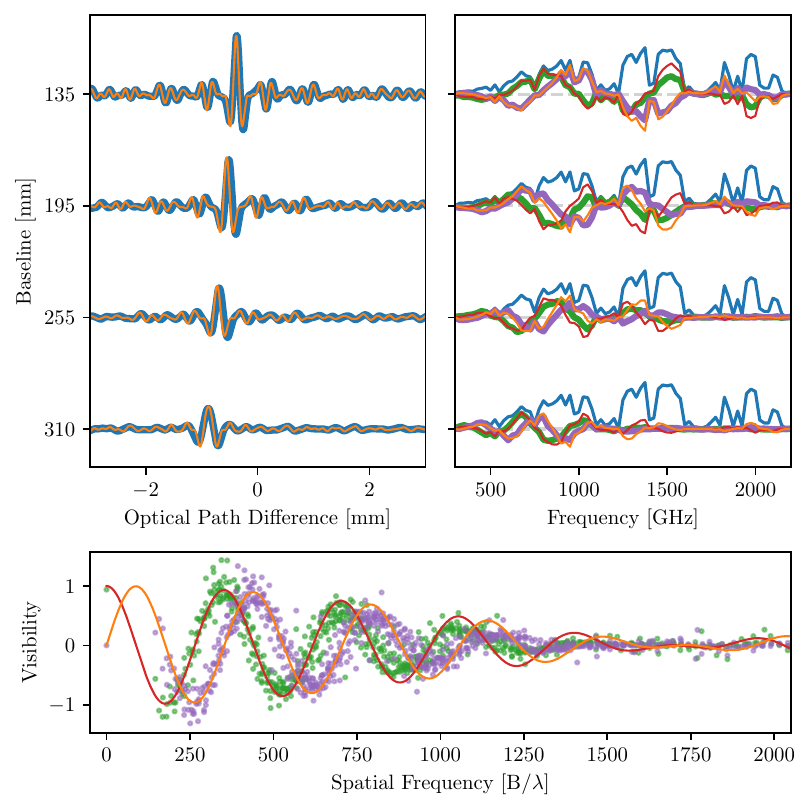}
	\includegraphics[width=0.8\linewidth]{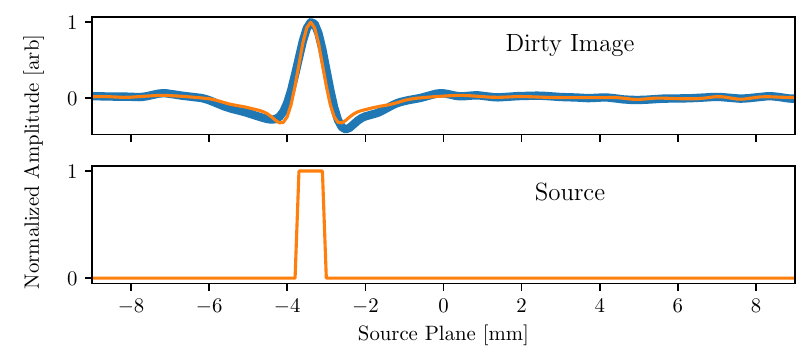}
	\caption[Offset 1 mm Slit Results]{Results for an offset $1\,\mathrm{mm}$ slit. Model is a $0.7\,\mathrm{mm}$ slit centered at $-3.4\,\mathrm{mm}$.}
	\label{fig:offsetslit}
\end{figure}

\begin{figure}
	\centering
	\includegraphics[width=0.8\linewidth]{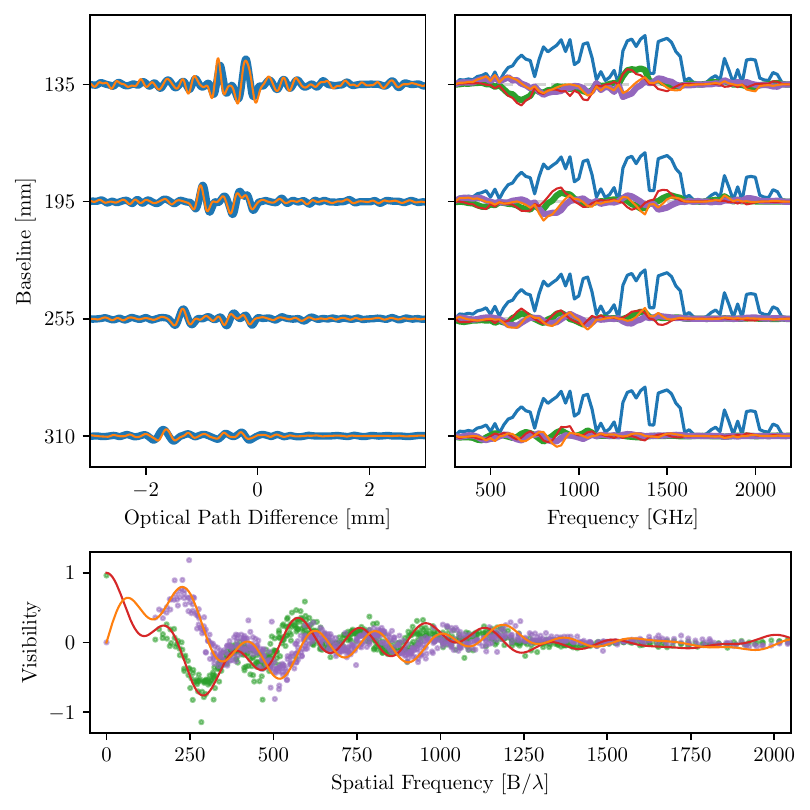}
	\includegraphics[width=0.8\linewidth]{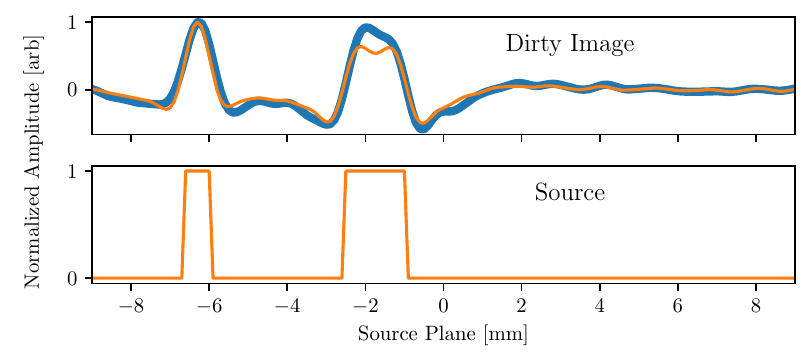}
	\caption[Double Slit Results]{Results for a $1\,\mathrm{mm}$ and $2\,\mathrm{mm}$ slit separated by $5\,\mathrm{mm}$.  Model is a $0.7\,\mathrm{mm}$ and $1.6\,\mathrm{mm}$ slit separated by $4.5\,\mathrm{mm}$.}
	\label{fig:doubleslit}
\end{figure}

Figs~\ref{fig:offsetslit} and \ref{fig:doubleslit} show the results when observing one slit (Fig.~\ref{fig:1mm_slit}) and two slits of different size (Fig.~\ref{fig:2slit_5mm}), respectively. In general, there is strong agreement between experimental results and theoretical predictions across all regimes of comparison, from interferogram to dirty image. Deviations from expectations typically manifest as a loss of coherence at high spatial frequencies. This is attributed to worse alignment of optical components at larger baselines and stronger requirements in terms of alignment and component surface quality for coherence with light at higher optical frequencies. Measurements of the OAP mirrors show significant striations reducing surface quality and increasing deviations from the ideal parabolic shape as one moves away from the optical axis of the collimating mirror \citep{scott:2024a}. 

It is emphasized that the geometry of the source for simulation purposes is similar but not equivalent to the physical source geometry. For example, the physical $1\,\mathrm{mm}$ slit has been modeled as a $0.7\,\mathrm{mm}$ slit (see Fig.~\ref{fig:offsetslit}). This reduced size results from  perspective or how the input apertures see the source through the collimating mirror. Consider placing your eye at the position of the static aperture in Fig.~\ref{fig:doubleFourier_diagram} and observing the source mask. You would not see the mask face-on but at an angle. This angle would reduce the perceived width of the slit and the thickness of the mask material would block even more of the slit. The apparent source size can be modeled as
\begin{equation}
    \label{eq:effective_source_size}
    s = w \cos(\theta) - t \tan(\theta) \;,
\end{equation}
where $\theta$ is the deviation from face-on viewing angle, $w$ is the physical width of the slit, and $t=2\,\mathrm{mm}$ is the thickness of the mask material. For the static aperture, $\theta \approx 8^\circ$, a slit width $w=1\,\mathrm{mm}$ evaluates to an apparent size of $s\approx0.7\,\mathrm{mm}$, consistent with the simulated source geometry. Since interference occurs for the parts of the image that overlap on the detector from the two arms of the interferometer, image reconstruction from interference effects is limited to the smallest apparent size of the source. 

It is also noted that the position of the slit mask with respect to the phase center of the instrument is not well constrained. The phase center on the source plane is only known to $\pm 2 \,\mathrm{mm}$ and the source mask is positioned manually with a similar precision. As such, we do not focus on the position of the source but rather the widths and relative positions of different source features. These results show that the DFI technique can accurately reconstruct source geometry and resolve features that cannot be distinguished using direct imaging (i.e., light from two slits integrated onto a single detector).

\subsubsection*{Spectral Dependence}

\begin{figure}[t]
	\centering
	\includegraphics[width=0.8\linewidth]{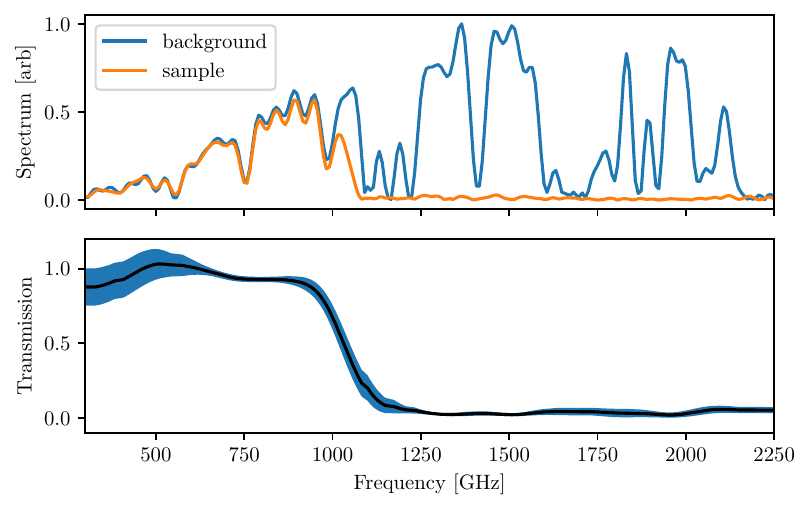}
	\caption[Low-pass Filter Transmission Curve]{Low-pass filter transmission curve. The top panel shows the full-pass spectrum (blue) obtained when viewing an unfiltered slit, backlit by a thermal source. The low-pass spectrum (orange) was obtained when viewing the slit masked by the low-pass filter material. Both spectra were obtained using FTS mode. The transmission curve in the lower panel (black) is computed as the ratio of the low-pass and the full-pass spectra. Uncertainty in transmission is conveyed by the blue shaded region.}
	\label{fig:transmissioncurve}
\end{figure}

In order to produce a source with spectral variation on the source geometry, we place a low-pass optical filter over the wide slit of the source mask shown in Fig.~\ref{fig:2slit_5mm}. The transmission curve of the low-pass filter is shown in Fig.~\ref{fig:transmissioncurve} which was obtained by comparing the spectra measured in FTS mode of a thermal source with and without the filter sample. The filter rapidly attenuates flux above the cutoff frequency of $\sim1.0\,\mathrm{THz}$. As a result, the source appears as two slits bellow the the cutoff frequency and one slit above the cutoff frequency. 

\begin{figure}[t]
	\centering
	\includegraphics[width=0.8\linewidth]{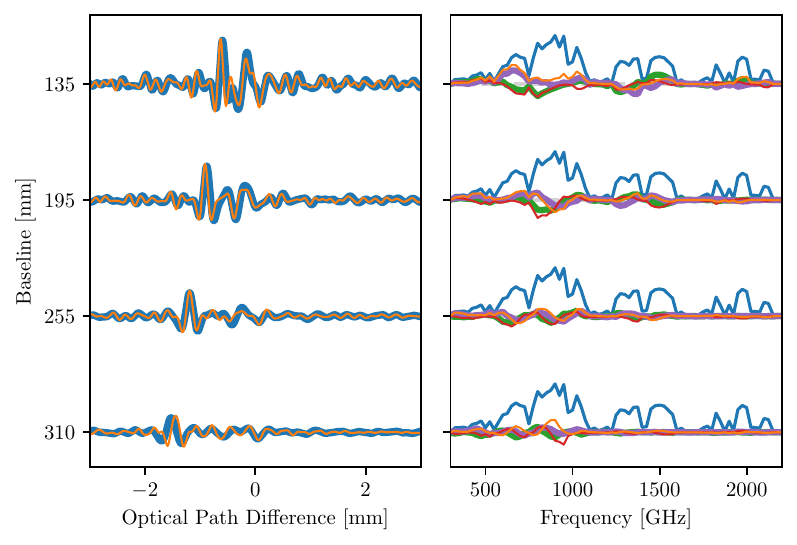}
	\caption[Full Band Nonuniform Spectrum Results]{Full bandwidth results for a $1\,\mathrm{mm}$ and $2\,\mathrm{mm}$ slit separated by $5\,\mathrm{mm}$, with the $2\,\mathrm{mm}$ slit masked by the $1\,\mathrm{THz}$ low-pass material. For modeling, the $2\,\mathrm{mm}$ slit is assigned the low-pass spectrum and the $1\,\mathrm{mm}$ slit is assigned the full-pass spectrum (see Fig.~\ref{fig:transmissioncurve}). Additional model details are described in Fig.~\ref{fig:lowerband} and \ref{fig:upperband}.}
	\label{fig:filteredvis}
\end{figure}

\begin{figure}
	\centering
	\includegraphics[width=0.7\textwidth]{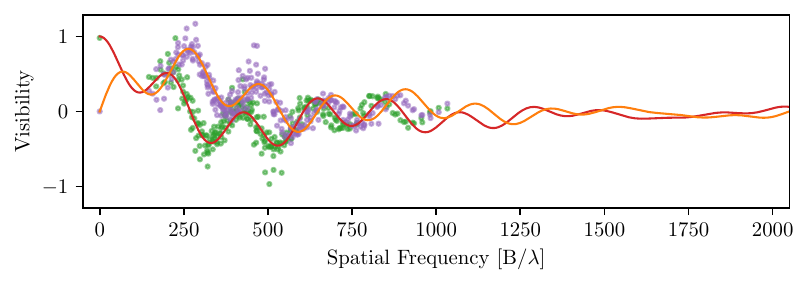}
	\includegraphics[width=0.7\textwidth]{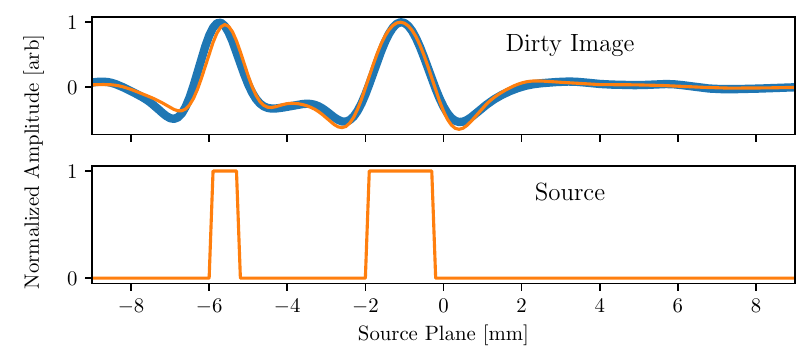}
	\caption[Lower Band Results]{\label{fig:lowerband}Lower band results. The model is a $1.6\,\mathrm{mm}$ slit and a $0.7\,\mathrm{mm}$ slit separated by $4.5\,\mathrm{mm}$.}
	\vfill
	\includegraphics[width=0.7\textwidth]{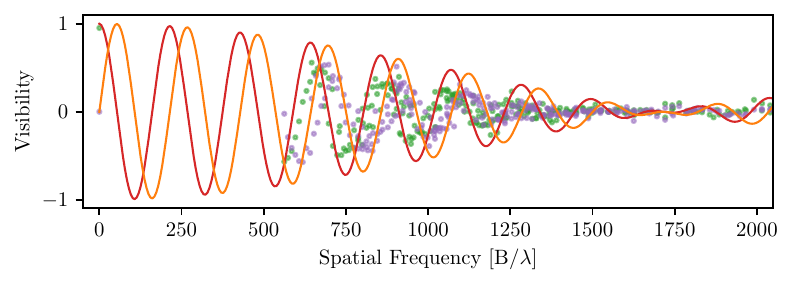}
	\includegraphics[width=0.7\textwidth]{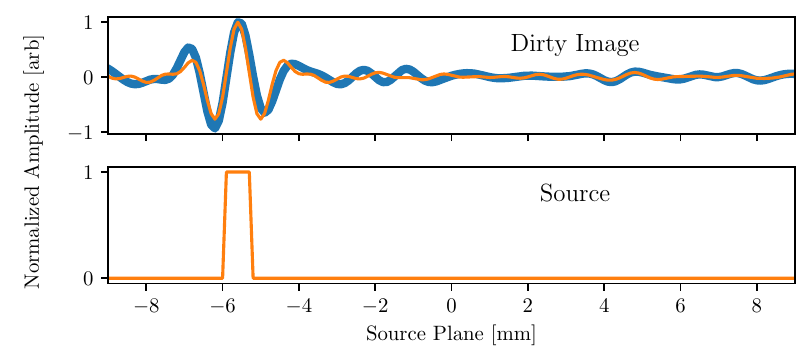}
	\caption[Upper Band Results]{\label{fig:upperband}Upper band results. The model is a $0.7\,\mathrm{mm}$ slit centered at $-5.6\,\mathrm{mm}$.}
\end{figure}

Data processing proceeds in a similar same way as described in \S~\ref{sec:processing}, however, in this case two $uv$-planes are constructed. One is populated with complex visibilities derived from frequencies below the cutoff frequency while the other plane is populated with samples above the cutoff frequency. Interferograms and complex spectra are shown in Fig.~\ref{fig:filteredvis} with complex visibility samples and image reconstruction for the lower and higher frequency samples shown in Figs.~\ref{fig:lowerband} and \ref{fig:upperband}, respectively. Again, we see strong agreement between experimental and simulated results, with the same loss of visibility at higher spatial frequencies and narrower slit widths previously discussed in \S~\ref{subsec:1d_sources}.

These results show the DFI technique is sensitive to the spectral variations in the source geometry. In this case, it was necessary to bin spectral samples below and above the low-pass filter cutoff frequency to increase $uv$ coverage. However, with more baseline coverage, this process can be applied to each spectral sample independently. 

\subsubsection*{Wide-Field}

The DFI is an imaging system. Like other imaging systems, the FOV of the instrument can be increased by extending the region of sensitivity on the detector plane, assuming the optical design is not limited by vignetting. This is typically achieved by introducing additional detector pixels, producing a detector array, where each pixel is sensitive to a different region of the source plane. Measurements from independent pixels can be combined to produce a wide-field image. 

\begin{figure}
	\centering
	\includegraphics[width=0.8\linewidth]{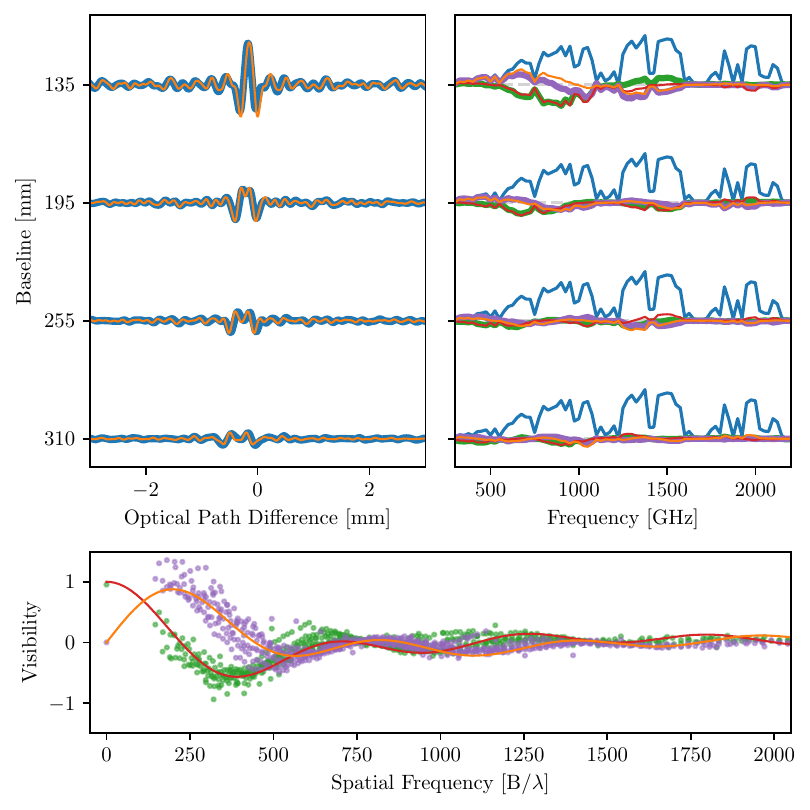}
	\includegraphics[width=0.8\linewidth]{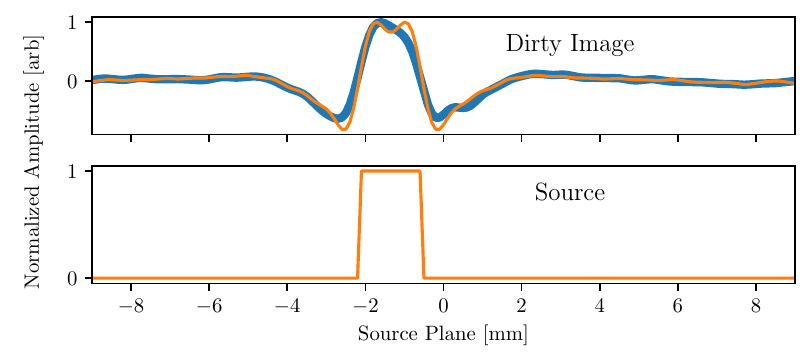}
	\caption[Wide Field A0 Results]{Wide field A0 results. Model is a $1.6\,\mathrm{mm}$ slit centered at $-1.4\,\mathrm{mm}$.}
	\label{fig:widea0vis}
\end{figure}

\begin{figure}
	\centering
	\includegraphics[width=0.8\linewidth]{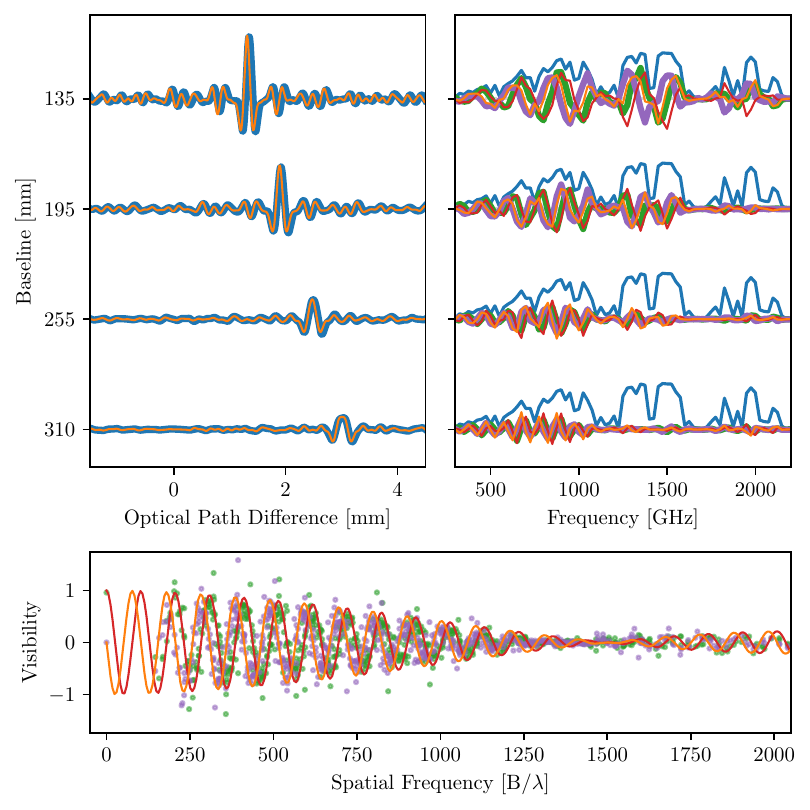}
	\includegraphics[width=0.8\linewidth]{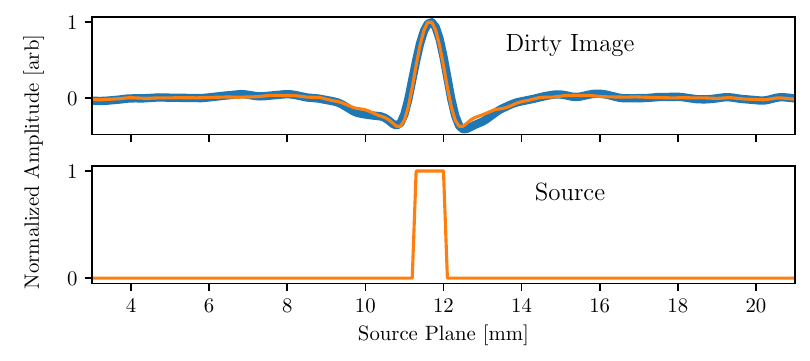}
	\caption[Wide Field A3 Results]{Wide field A3 results. Model is a $0.7\,\mathrm{mm}$ slit centered at $11.6\,\mathrm{mm}$.}
	\label{fig:widea3vis}
\end{figure}

Wide-field capabilities are demonstrated using the source mask shown in Fig.~\ref{fig:2slit_12mm}. The source mask has two slits separated by a distance that a single detector could not measure both simultaneously. The source mask is placed on the source plane such that the individual slits are roughly centered in the FOV of the adjacent A0 and A3 detectors. We perform an observation as previously described and process the data from both detectors in an identical way. Results for detector A0 and A3 are shown in Figs.~\ref{fig:widea0vis} and \ref{fig:widea3vis}, respectively. Again, experimental and simulated results show significant agreement. 


The results are consistent with a $0.7\,\mathrm{mm}$ and a $1.6\,\mathrm{mm}$ slits separated by $13\,\mathrm{mm}$, which is compared to the source mask with $1.0\,\mathrm{mm}$ and $2.0\,\mathrm{mm}$ slits separated by $12\,\mathrm{mm}$. The reduced slit width is commensurate with equation~\ref{eq:effective_source_size}, but the increased separation between slits is unexpected. There is a well known obliquity correction for imaging FTS \citep{davis:2001a}, which would apply to a DFI where the optical path difference $z$ is modified to
\begin{equation}
    \label{eq:obliquity_correction}
    z' = z \cos(\alpha) \;,
\end{equation}
where $\alpha$ is the angle from the optical axis to the off-axis source. However, in this case the cosine factor is on the order of $10^{-5}$ and cannot explain the discrepancy. 

We do not expect that detection specific corrections are required to account for off-axis sources, apart from equation~\ref{eq:obliquity_correction}. Indeed, interferograms identical in shape are measured when a slit is placed in the small region of overlap between detectors A0 and A3 \citep{scott:2024a}. This is also the case for detectors A0 and C2 (the detector on the other side of A0; see Fig.~\ref{fig:2slit_12mm}). Repeating the phase center calibration procedure described in \S~\ref{sec:processing} resulted in different OPD offsets: $0.15\pm0.03 \,\mathrm{mm}$ for the A0/A3 overlap region; $0.28\pm0.03 \,\mathrm{mm}$ for the center of A0; $0.42\pm0.02 \,\mathrm{mm}$ for the A0/C2 overlap region. Note that OPD corrections were the same for both detectors in cases where the slit was placed in an overlapping region of two detectors. It is concluded that no detector specific correction is needed, but that the phase center depends on position in the source plane. In effect, the source plane is distorted such that the separation between sources on the A3 side of the optical axis are increased and the separation on the C2 side are decreases. These distortions are consistent with the coma aberrations associated with the dual OAP focal system used for the DFI testbed \citep{scott:2024a}.

\subsection{Two-Dimensional Sources}
\label{subsec:2d_sources}

A two-dimensional source is achieved using the mask shown in Fig.~\ref{fig:2d_mask}. Unlike previous targets, this source has spatial variation along multiple axes, and these variations cannot be effectively measured with a single baseline orientation; the baseline vector must be rotated to sample a two-dimensional $uv$-space. In practice, a real space-based observatory would rotate the physical baseline. In the case of the testbed instrument, the source mask is rotated and the relationship between the source rotation angle and the baseline vector rotation angle is $\theta_{src} = -\theta_{\vec{B}}$, which is relevant for $uv$-plane projection of complex visibility samples (see equation~\ref{eq:uv_projection}). 

\begin{figure}
	\centering
	\includegraphics[width=0.8\linewidth]{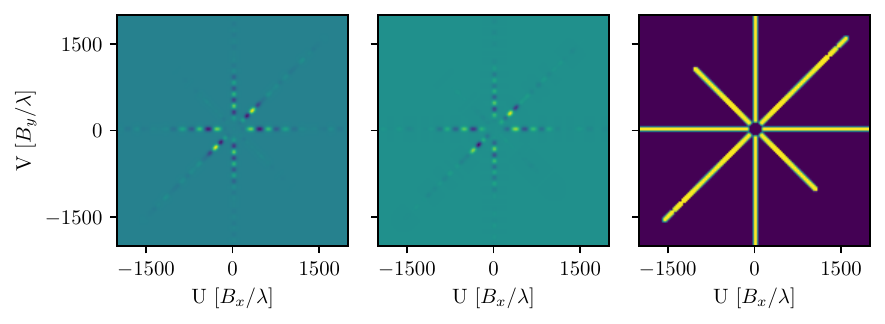}
	\caption[Two Dimensional $uv$ Sampling]{Two dimensional $uv$ sampling. The left and center panels show the measured real and complex $uv$ samples after regridding, respectively. The right panel shows the $uv$-plane coverage for this experiment.}
	\label{fig:uvsamples}
\end{figure}

Due to the geometry of the source mask, most of the signal in the $uv$-plane is distributed along spatial frequencies with $\theta_{\vec{B}} = \{0^\circ, 45^\circ, 90^\circ \}$. We focus on these baseline orientations for efficiency but also sample $\theta_{\vec{B}} = 135^\circ$ to confirm the lack of signal. Observations and data processing proceeded as described in \S~\ref{sec:processing}. The only notable additions from previous observations are that the source was rotated and that the corresponding baseline rotation was used to project complex visibility samples into a two-dimensional $uv$-plane. The resulting complex visibilities after regridding are shown in Fig.~\ref{fig:uvsamples}. Also included in this figure are the regions on the $uv$-plane that were sampled. We note that there is little signal along the $\theta_{\vec{B}} = 135^\circ$ direction, as expected. 

\begin{figure}
	\centering
	\includegraphics[width=0.8\linewidth]{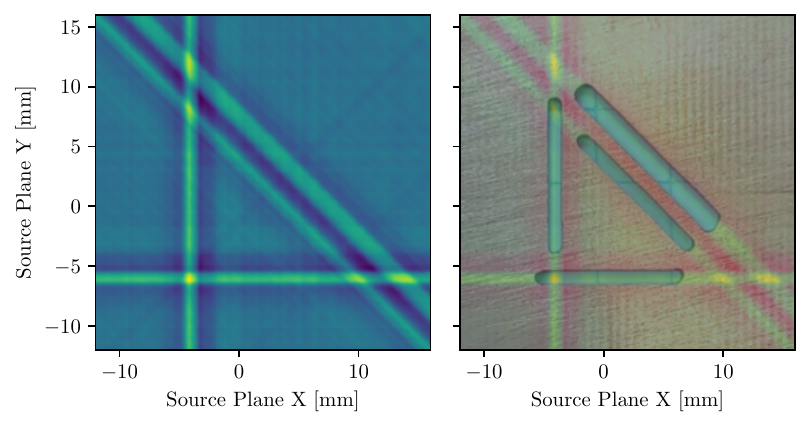}
	\caption[Two Dimensional Image Reconstruction]{Two dimensional image reconstruction. The left panel shows the resulting dirty image, which is overlayed onto a picture of the physical slit mask used for this experiment in the right panel. Although the lengths of the slits are not evident in the dirty image, the widths, orientations, and relative positions are well represented by the dirty image.}
	\label{fig:sourcerec}
\end{figure}

The reconstructed dirty image is shown in Fig.~\ref{fig:sourcerec}. In this figure, the dirty image is also overlayed with the source mask to show the agreement between the image reconstruction and physical source. It can be seen that the widths, orientations, and relative positions of the slits are in good agreement. The lengths of the slits are not well represented, but this could be improved in principal with more extended sampling of the $uv$-plane. However, the required low amplitude high spatial frequencies would be difficult to measure with this instrument given the limited SNR. In addition to the previously discussed issues with sampling high spatial frequencies, in this case, the source rotation would need to be known to high precision to accurately project complex visibility samples into the $uv$-plane far away from the origin. 

A common heuristic for interferometry is that the surface quality of optics must have a root-mean-squared deviation $\le \lambda/10$ in order for the light to remain coherent. Using a similar idea, the rotational precision of the source should satisfy $r \sigma_{\theta_{src}} \le \lambda/10$. That is, a point source located a distance $r$ from the optical axis on the source plane must be within $\lambda/10$ of its expected position. In the representative case for this observation where $r\approx 10\,\mathrm{mm}$ and $\nu = 2.2 \,\mathrm{THz}$, the rotational precession required is $\sigma_{\theta_{src}} \le 0.1^\circ$. This is much more demanding than the $\pm3^\circ$ precision obtainable by our manual mounting and rotation strategy. 

\section{Conclusion}
\label{sec:conclusion}

The FIR is an underutilized band in astronomy. Due to its abundance of important line and continuum emission features it has the potential to be one of the most productive bands for studying star formation, proto-planetary disks, life sustaining planets, the coevolution of AGN and stellar activity in galaxies, and more. The science deficit in this area is largely due to limited spatial resolution.  

A promising instrument concept that preserves the rich spectral information of the FIR while delivering the necessary spatial resolution is the DFI, an instrument that combines a spatial interferometer and an FTS to provide spectral images with enhanced spatial resolution. FTS has been used to great success in past FIR observatories, and the DFI has been put forward as a concept mission by both the NASA and ESA communities. A handful of testbed instruments have been studied to validate the DFI, yet none have fully demonstrated the capabilities of this observing technique. This work provides independent demonstrations of all the DFI capabilities in the FIR. 

In this work, we have shown how the DFI can provide enhanced spatial resolution, resolving features that cannot be distinguished by individual detectors. We have shown that the DFI is sensitive to the spectral variations of the source geometry; the source has a different shape at different optical frequencies. We have shown that the FOV of the instrument can be increased by combining the data from independent detectors sensitive to different regions of the source plane. Finally, we have shown that two-dimensional sources can be reconstructed using different baseline orientations. 

In the course of these demonstrations, we have derived a general instrument response function and expanded on how the measured signal relates to the source being observed. Instrumental effects were identified which were used to develop detailed calibration and processing procedures.

In general, experimental and simulated results have shown significant agreement. The modest deviations include underestimating the source size, loss of visibility at high spatial frequencies, and image distortions in the wide-field case. The source size discrepancy is attributed to perspective where the apertures see the source through the collimating mirror at an angle reducing the apparent source size. The loss of visibility is due to the stronger requirements to maintain coherence at higher optical frequencies in addition to imperfect alignment partially due to the imperfect collimating mirror shape. The wide-field distortions are attributed to off-axis aberrations associated with dual OAP optical system. 

By successfully demonstrating these capabilities independently, this research validates the DFI as a mature and promising architecture for the next generation of FIR observatories. Ultimately, the successful implementation of DFI technology will be essential for resolving the fundamental physical processes of star and planet formation across cosmic time.

\begin{acknowledgments}

The authors acknowledge funding and infrastructure support from Canada Foundation for Innovation (CFI), Alberta Research Capacity Program (RCP), the Natural Science and Engineering Research Council of Canada (NSERC), Canada Research Chairs (CRC), the University of Lethbridge, CMC Microsystems, and Canada’s National Design Network (CNDN). In addition, the authors would like to acknowledge Mr. Adam Sundberg, Mr. Geoff Sitwell, Ms. Megan Ritchie, Mr. Klondike Taylor, Mr. Rodderick MacCrimmon, and Mr. Braden Gail for their contributions to laboratory infrastructure which made this work possible.

\end{acknowledgments}

\begin{contribution}

\noindent Jeremy P. Scott: Detailed instrument design, alignment, hardware control, and data acquisition software. Conducted the experiments, performed instrument characterization, calibrated the data and developed the data processing pipeline, and served as the primary author of the manuscript.  

\noindent Locke D. Spencer: Contributed to high-level instrument design, detector software development, project funding acquisition, and manuscript review and editing.  

\noindent Chris Benson: Assisted with instrument alignment and conducted detector characterization.  

\noindent David Naylor: Provided the necessary hardware for instrument characterization, including the thermal source.  

\noindent Berke Ricketti: Contributed to the overall project collaboration.  

\noindent Dinula De Silva: Contributed to the development of the data acquisition software.  

\noindent Vince Weiler: Developed the linear translation stage control software and provided the source masks for the experimental scenes. 


\end{contribution}

%

\software{
    NumPy \citep{harris:2020a},
    SciPy \citep{virtanen:2020a},
    Matplotlib \citep{Hunter:2007a}
}




\bibliography{full_refs}{}
\bibliographystyle{aasjournalv7}



\end{document}